\documentclass[doublecol]{epl2}
\usepackage{bm}
\usepackage{graphicx}
\bibstyle{approve.bib}
\usepackage{amssymb}
\usepackage{amsmath}
\usepackage{esint}
\usepackage{epstopdf}
\usepackage{color}
\usepackage{gensymb}
\usepackage{mathtools}
\usepackage{suffix}

\newcommand{\be}{\begin{equation}}
\newcommand{\m}{\mbox}
\newcommand{\ma}{\mathrm}
\newcommand{\ph}{\rm pH}
\newcommand{\ra}{\xrightleftharpoons}
\newcommand{\ee}{\end{equation}}
\newcommand{\bea}{\begin{eqnarray}}
\newcommand{\eea}{\end{eqnarray}}

\newcommand{\bE}{\mathbf{E}}

\newcommand{\lm}{L_-}

\newcommand{\e}{\varepsilon}

\newcommand{\ld}{\lambda_{\rm d}}
\newcommand{\lb}{\lambda_{\rm b}}

\begin{document}

\title{pH-mediated regulation of polymer transport through SiN pores}

\author{Sahin Buyukdagli\inst{1,2}\footnote{email:~\texttt{buyukdagli@fen.bilkent.edu.tr}} and T. Ala-Nissila\inst{3,4}\footnote{email:~\texttt{Tapio.Ala-Nissila@aalto.fi}}}

\institute{
  \inst{1} Department of Physics, Bilkent University, Ankara 06800, Turkey\\
  \inst{2} QTF Centre of Excellence, Department of Applied Physics, Aalto University, FI-00076 Aalto, Finland.\\
  \inst{3} Department of Applied Physics and QTF Center of Excellence, Aalto University School of Science, P.O. Box 11000, FI-00076 Aalto, Espoo, Finland\\
  \inst{4} Interdisciplinary Centre for Mathematical Modelling and Department of Mathematical Sciences, Loughborough University, Loughborough, Leicestershire LE11 3TU, United Kingdom
}

\abstract{We characterize the $\ph$ controlled polymer capture and transport thorough silicon nitride (SiN) pores subject to protonation. 
A charge regulation model able to reproduce the experimental zeta potential of SiN pores is coupled with electrohydrodynamic 
polymer transport equations. The formalism can quantitatively explain the experimentally observed non-monotonic $\ph$ dependence of avidin 
conductivity in terms of the interplay between the electroosmotic and electrophoretic drag forces on the protein. 
We also scrutinize the DNA conductivity of SiN pores. We show that in the low $\ph$ regime where the amphoteric pore is cationic, 
DNA-pore attraction acts as an electrostatic trap. This provides a favorable condition for 
fast polymer capture and extended translocation required for accurate polymer sequencing.}


\maketitle

\section{Introduction}

The rapid progress in biotechnological applications requires an increasingly high degree of precision in bioanalytical approaches such as polymer translocation~\cite{e1,e2,e3,e4,e5}. 
Accurate control over the mobility of confined polymers is vital for improving the sensitivity of this biosequencing technique~\cite{muthu}. 
Over the last two decades, this technological requirement has motivated intense research into the characterization of entropic~\cite{Saka1,n2}  and electrohydrodynamic effects~\cite{aks3,Ghosal,mutalp,Buy2014,Buy2017,Buy2018}  on polymer translocation. 

In driven polymer transport through amphoteric silicon nitride (SiN) pores subject to protonation, the acidity of the buffer solution is a critical control factor enabling the radical alteration 
of the forces driving the polymer mobility. More precisely, the inversion of the pore surface charge upon $\ph$ tuning can reverse the direction of the 
electro-osmotic (EO) flow drag~\cite{Firnkes2010} and also switch the nature of polymer-membrane interactions between repulsive and attractive~\cite{muthuNANO}. 
The quantitatively accurate characterization of this mechanism can thus provide an efficient control of the polymer translocation dynamics.

Previous charge regulation theories have ingeniously characterized the effect of surface protonation on macromolecular interactions~\cite{Rudi1,Rudi2,Netz1,Netz2}. 
However, a polymer translocation model able to account for the $\ph$ controlled alteration of the pore electrohydrodynamics and polymer-pore interactions is still missing. 
In this Letter, we develop such a polymer translocation model. Within our formalism, we first explain the experimentally measured non-monotonic $\ph$ dependence of avidin translocation 
rates in terms of the electrohydrodynamic forces acting on the avidin protein of amphoteric nature~\cite{Firnkes2010}. Then, we investigate the dsDNA conductivity of 
SiN pores and shed light on an electrostatic polymer trapping mechanism allowing favorable conditions for fast polymer capture and slow translocation required for accurate biosequencing
and related applications. 

\section{Theory}

\subsection{Polymer transport model}

\begin{figure}
\includegraphics[width=1\linewidth]{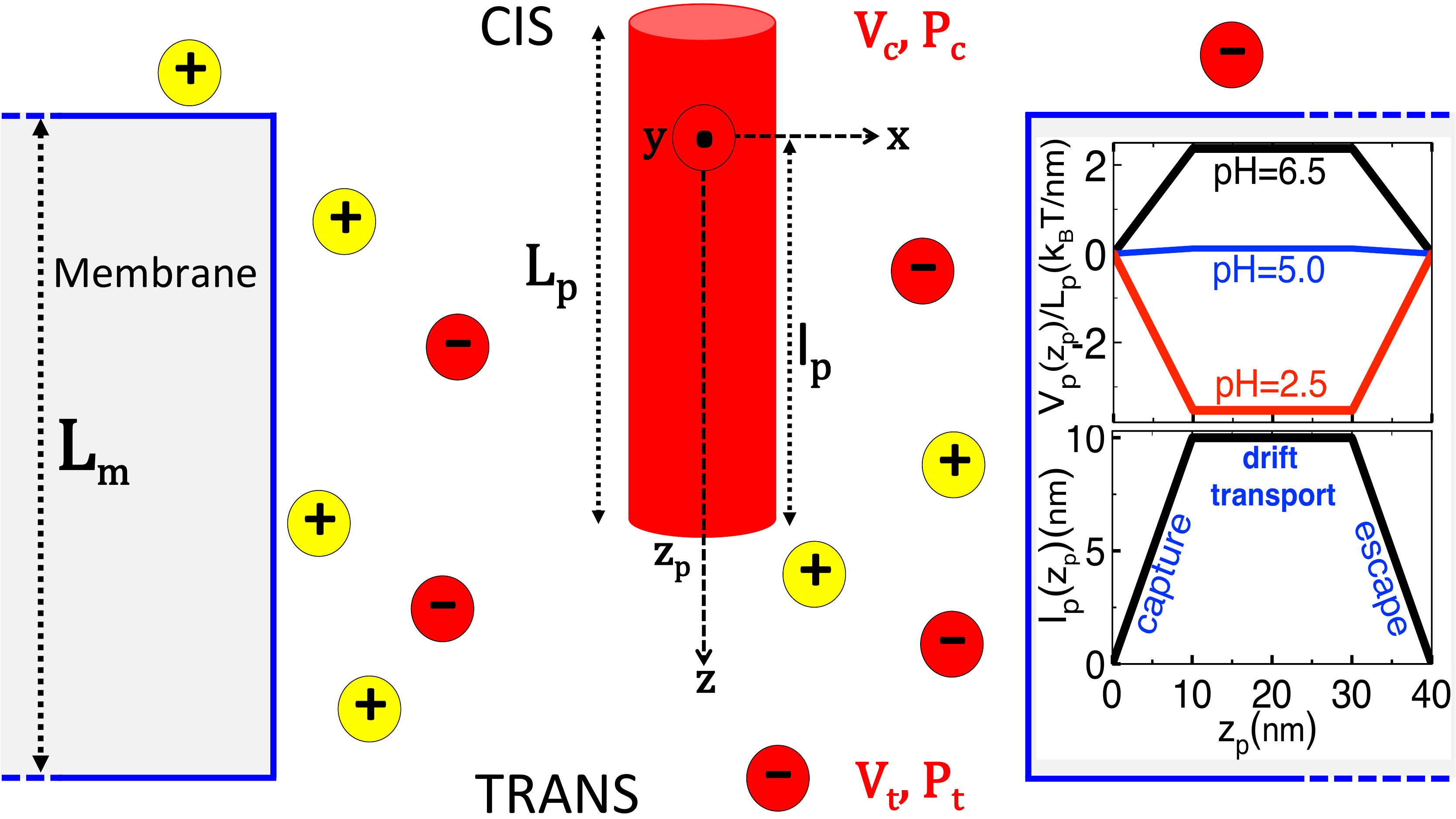}
\caption{(Color online) Schematic depiction of the polyelectrolyte translocating along the $z$ axis of the pore confining the KCl solution of bulk concentration $\rho_{\rm b}$. 
The polyelectrolyte is a cylinder of radius $a$, total length $L_{\rm p}$, and its portion located in the pore is $l_{\rm p}$. The pore is a cylinder of radius $d$ and length $L_{\rm m}$, \textcolor{black}{connecting the cis and trans sides of the membrane.} 
Charge transport through the pore takes place under the effect of the voltage $\Delta V=V_{\rm t} - V_{\rm c}$ resulting in the field $\bE=-\Delta V/L_{\rm m}\hat{u}_z$ and 
pressure gradient $\Delta P=P_{\rm c} - P_{\rm t}$. \textcolor{black}{The inset displays the polymer-membrane interaction potential $V_p(z_p)$ and the length $l_p(z_p)$ in Eq.~(\ref{lpzp}) for the parameter values of Fig.~\ref{fig4}(c).}}
\label{fig1}
\end{figure}

We briefly review here the polymer translocation model initially developed in Ref.~\cite{Buy2017} for fixed surface charge conditions. 
The model is depicted in Fig.~\ref{fig1}. The nanopore is a cylindrical hole embedded in a SiN membrane of surface charge density $\sigma_{\rm m}$. 
In this work, the pore radius and length will be fixed to the experimental values of $d=10$ nm and $L_{\rm m}=30$ nm of Ref.~\cite{Firnkes2010}. The pore is in contact with an ion reservoir confining the 
KCl solution of density $\rho_{\rm b}$. The polymer translocates along the $z$ axis is a rigid cylinder of radius $a=1$ nm, length $L_{\rm p}$, and surface charge density $\sigma_{\rm p}$. 
The charge transport through the pore is driven by the externally applied hydrostatic pressure $\Delta P$ and voltage $\Delta V$. 

The coordinate of the polymer end $z_{\rm p}$  is chosen as the reaction coordinate of the translocation while 
$l_{\rm p}$ is the length of the polymer portion in the pore. The translocation dynamics is characterized by the diffusion equation
\bea\label{con}
\partial_tc(z_{\rm p},t)&=&-\partial_{z_{\rm p}}J(z_{\rm p},t);\\
\label{cur}
J(z_{\rm p},t)&=&-D_{\rm\textcolor{black}{b}}\partial_{z_{\rm p}}c(z_{\rm p},t)+v_{\rm p}(z_{\rm p})c(z_{\rm p},t),
\eea
where $c(z_{\rm p},t)$ is the density and $J(z_{\rm p},t)$ the flux of the translocating polymer, with the \textcolor{black}{bulk} diffusion coefficient 
$D_{\rm\textcolor{black}{b}}=\ln(L_{\rm p}/2a)/(3\pi\eta L_{\rm p}\beta)$ including the inverse thermal energy $\beta=1/(k_{\rm B}T)$ and solvent viscosity $\eta=8.91\times10^{-4}$ Pa s~\textcolor{black}{\cite{cyl1,cyl2}}. 
In Eq.~(\ref{cur}), the first term is Fick's law \textcolor{black}{accounting for the diffusion-driven polymer dynamics. The second convective flux term originates from the polymer velocity $v_{\rm p}(z_{\rm p})$ induced by external electrohydrodynamic forces.} In Ref.~\cite{Buy2017}, from the coupled solution of the Stokes and Poisson-Boltzmann (PB) equations, the liquid velocity $u_{\rm c}(r)$ and polymer velocity $v_p(z_p)$ satisfying the 
no-slip conditions $u_{\rm c}(d)=0$ and $u_{\rm c}(a)=v_{\rm p}(z_{\rm p})$ were derived as
\bea\label{vels}
u_{\rm c}(r)&=&\mu E\left[\phi(d)-\phi(r)\right]-\beta D_{\rm p}(r)\frac{\partial V_{\rm p}(z_{\rm p})}{\partial z_{\rm p}}\nonumber\\
&&+\frac{\Delta P}{4\eta L_{\rm m}}\left[d^2-r^2-2a^2\ln\left(\frac{d}{r}\right)\right];\\
\label{velp}
v_{\rm p}(z_{\rm p})&=&v_{\rm dr}-\beta D_{\rm p}(a)\frac{\partial V_{\rm p}(z_{\rm p})}{\partial z_{\rm p}}.
\eea
The first term of Eq.~(\ref{velp}) corresponds to the drift velocity induced by the the voltage and the pressure gradient,
\be
\label{vdr}
v_{\rm dr}=-\frac{\mu\Delta V}{L_{\rm m}}\left[\phi(a)-\phi(d)\right]+\frac{\gamma a^2\Delta P}{4\eta L_{\rm m}},
\ee
with the electrophoretic (EP) mobility coefficient $\mu=\e_{\rm w} k_{\rm B}T/(e\eta)$ including the electron charge $e$ \textcolor{black}{and solvent permittivity $\e_{\rm w}=80$}, the geometric coefficient 
$\gamma=d^2/a^2-1-2\ln\left(d/a\right)$, and the electrostatic potential $\phi(r)$ induced by the polymer and membrane charges. \textcolor{black}{In the bracket of
Eq.~(\ref{vdr}), the zeta potential terms $\phi(a)$ and $-\phi(d)$ correspond respectively to the contribution from the EP and EO drag forces to the polymer velocity}. Then, the second term of Eq.~(\ref{velp})  includes the pore diffusion coefficient 
$D_{\rm p}(r)=\ln(d/r)/(2\pi\eta L_{\rm p}\beta)$, and the electrostatic coupling potential between the polymer and membrane charges 
$\beta V_{\rm p}(z_{\rm p})=\psi_{\rm p} l_{\rm p}(z_{\rm p})$. This potential includes the energy density
\be\label{enden}
\psi_{\rm p}=2\pi a\sigma_{\rm p}\phi_{\rm m}(a),
\ee
with the polymer potential induced solely by the membrane charges $\phi_{\rm m}(r)\equiv\lim_{\sigma_{\rm p}\to0}\phi(r)$, 
and the position-dependent length of the polymer portion in the pore
\bea
\label{lpzp}
l_{\rm p}(z_{\rm p})&=&z_{\rm p}\theta(L_- - z_{\rm p}) + L_- \theta(z_{\rm p} - L_-)\theta(L_+ - z_{\rm p})\nonumber\\
&&+ (L_{\rm p} + L_{\rm m} - z_{\rm p})\theta(z_{\rm p} - L_+),
\eea
\textcolor{black}{where we defined} the auxiliary lengths $L_- = \mathrm{min}(L_{\rm m},L_{\rm p})$ and $L_+ = \mathrm{max}(L_{\rm m},L_{\rm p})$. The terms on the r.h.s. of Eq.~(\ref{lpzp}) 
correspond respectively to the regimes of polymer capture, transport at drift velocity $v_{\rm dr}$, and escape from the pore \textcolor{black}{(see the bottom plot in the inset of Fig.~\ref{fig1})}.

The polymer translocation rate follows from the steady-state solution of Eqs.~(\ref{con})--(\ref{cur}) characterized by a uniform flux $J(z_{\rm p},t)=J_0$ , 
with the fixed density condition at the pore entrance $c(z_{\rm p}=0)=c_{\rm cis}$ and an absorbing boundary at the pore exit $c(z_{\rm p}=L_{\rm p}+L_{\rm m})=0$. 
The translocation rate defined as $R_{\rm p} \equiv J_0/c_{\rm cis}$ reads~\cite{Buy2017}
\be
\label{rc}
R_{\rm p}=\frac{D_{\rm\textcolor{black}{b}}}{\int_0^{L_{\rm m} + L_{\rm p}}\textcolor{black}{\mathrm{d}z_p}e^{\beta U_{\rm p}(z_{\rm p})}},
\ee
with the effective polymer potential
\be
\label{polp}
U_{\rm p}(z_{\rm p})=\frac{D_{\rm p}(a)}{D_{\rm\textcolor{black}{b}}}V_{\rm p}(z_{\rm p})-\frac{v_{\rm dr}}{\beta D_{\rm\textcolor{black}{b}}}z_{\rm p}.
\ee
Defining the characteristic inverse lengths embodying the effect of the drift~(\ref{vdr}) and the barrier~(\ref{enden}),
\be
\label{len}
\ld=\frac{v_{\rm dr}}{D_{\rm\textcolor{black}{b}}};\hspace{5mm}\lb=2\pi a\sigma_{\rm p} \phi_{\rm m}(a)\frac{D_{\rm p}(a)}{D_{\rm\textcolor{black}{b}}},
\ee
the effective polymer potential~(\ref{polp}) can be expressed as
\be
\label{polp2}
\beta U_{\rm p}(z_{\rm p})=\lb l_{\rm p}(z_{\rm p})-\ld z_{\rm p}.
\ee
\textcolor{black}{The analytical expression for $R_{\rm p}$ obtained from Eqs.~(\ref{rc}) and~(\ref{polp2}) can be found in Ref.~\cite{Buy2017}.} We finally note that in the \textit{drift regime} $\ld\gg\lb$ where polymer-pore interactions are negligible $V_{\rm p}(z_{\rm p})\ll k_{\rm B}T$, Eq.~(\ref{rc}) yields the drift-driven polymer transport behavior $R_{\rm p} \approx v_{\rm dr}$.


The polymer translocation time defined as the mean first passage time between the cis and trans sides is 
\be\label{tautot}
\tau_p=\tau_c+\tau_d+\tau_e,
\ee
with the time of polymer capture $\tau_c=I(0,L_-)$, diffusion $\tau_d=I(L_-,L_+)$, and escape $\tau_e=I(L_+,L_p+L_m)$, where we defined the auxiliary integral $I(z_i,z_f)=D^{-1}\int_{z_i}^{z_f}\mathrm{d}z'e^{\beta U_p(z')}\int_{0}^{z'}\mathrm{d}z''e^{-\beta U_p(z'')}$~\cite{muthu,Buy2017}. The analytical form of the translocation time can be also found in Ref.~\cite{Buy2017}. In the drift regime $\ld\gg\lb$, the translocation time reduces to its drift limit $\tau_{\rm p}\approx\tau_{\rm dr}=(L_{\rm p}+L_{\rm m})/v_{\rm dr}$.

\subsection{Charge regulation model}

Here, we derive the pH dependent surface charge density of the SiN pore. To this end, within the framework of the chemical reaction scheme proposed in Ref.~\cite{Harame1987}, 
we will extend the charge regulation model of Ref.~\cite{Hoog2009} to include the positively charged amine groups. The surface of the SiN pore is composed of amphoteric silanol (SiOH) 
and primary amine ($\mbox{SiNH}_2$) groups. The hydrolysis reactions resulting in SiN
\bea
\label{1}
&&\mbox{Si}_3\m{N}+\m{H}_2\m{O}\rightarrow\m{Si}_2\m{NH}+\m{SiOH};\\
\label{2}
&&\mbox{Si}_2\m{NH}+\m{H}_2\m{O}\rightarrow\m{Si}\m{NH}_2+\m{SiOH}
\eea
imply that on the pore surface, there are two silanol groups for every primary amine group~\cite{Harame1987}. Thus, the number of amphoteric groups $N_{\rm a}$ and primary amine 
sites $N_{\rm p}$ are related as $N_{\rm a}=2N_{\rm p}$. In the following, we assume that the amphoteric and primary amine groups are 
characterized by the same surface number density $\sigma_{\rm 0m}=(N_{\rm a}+N_{\rm p})/S=3N_{\rm p}/S$, with the area of the cylindrical pore $S=2\pi dL_{\rm m}$.

The reactions for the silanol groups on the pore are
\be\label{4}
\m{SiOH}\ra[k_{\rm r}]{k_{\rm d}}\m{SiO}^-+\m{H}^+;\;\m{SiOH}+\m{H}^+\ra[l_{\rm d}]{l_{\rm r}}\m{SiOH}_2^+,
\ee
with the corresponding mass action laws
\bea\label{6}
K_{\rm m}&=&10^{-\ma{pK}_{\rm m}}=\frac{k_{\rm d}}{k_{\rm r}}=\frac{\left[\m{SiO}^-\right]\left[\m{H}^+\right]}{\left[\m{SiOH}\right]};\\
\label{7}
L_{\rm m}&=&10^{-\ma{pL}_{\rm m}}=\frac{l_{\rm d}}{l_{\rm r}}=\frac{\left[\m{SiOH}\right]\left[\m{H}^+\right]}{\left[\m{SiOH}_2^+\right]},
\eea 
where $K_{\rm m}$ and $L_{\rm m}$ are the dissociation rates. Then, the $\m{H}^+$ binding reaction for primary amine groups is
\be\label{8}
\m{SiNH}_2+\m{H}^+\ra[t_{\rm d}]{t_{\rm r}}\m{SiNH}_3^+,
\ee
with the reaction rate $T_{\rm m}$ defining the mass action law
\be\label{9}
T_{\rm m} =10^{-\ma{pT}_{\rm m}}=\frac{t_{\rm d}}{t_{\rm r}}=\frac{\left[\m{SiNH}_2\right]\left[\m{H}^+\right]}{\left[\m{SiNH}_3^+\right]}.
\ee
In Eqs.~(\ref{6})--(\ref{7}) and~(\ref{9}), the $\m{H}^+$ density on the pore surface is given by $\left[\m{H}^+\right]=\left[\m{H}^+\right]_{\rm b} e^{-\phi(d)}$ 
where the $\m{H}^+$ density in the bulk reservoir is related to the acidity of the solution as $\ma{pH}=-\log_{10}\left\{\left[\m{H}^+\right]_{\rm b}\right\}$.

In order to derive the pore surface charge density $\sigma_{\rm m}$, we express first the density of the chemical species on the pore surface in terms of their rates 
$\alpha$, $\beta$, and $\gamma$ as
\be\label{10} 
\left[\m{SiO}^-\right]=N_{\rm a}\alpha;\;\left[\m{SiOH}_2^+\right]=N_{\rm a}\beta;\;\left[\m{SiOH}\right]=(1-\alpha-\beta)N_{\rm a},
\ee
for the amphoteric surface groups and
\be\label{110}
\left[\m{SiNH}_3^+\right]=N_{\rm p}\gamma;\hspace{5mm}\left[\m{SiNH}_2\right]=N_{\rm p}(1-\gamma)
\ee
for the primitive amine groups. Noting that the net surface charge is $Q_{\rm net}=S\sigma_{\rm m}=(\beta-\alpha)N_{\rm a}+\gamma N_{\rm p}$, the 
pore surface charge density $\sigma_{\rm m}=Q_{\rm net}/S$ follows in the form $\sigma_{\rm m}=\sigma_{\rm 0m}\left[2(\beta-\alpha)+\gamma\right]/3$. 
Calculating the rates $\alpha$ and $\beta$ from the solution of Eqs.~(\ref{6})--(\ref{7}) and~(\ref{10}), and the rate $\gamma$ from Eqs.~(\ref{9}) and~(\ref{110}), one finally obtains
\bea\label{11}
\sigma_{\rm m}&=&\left\{\frac{2\left[10^{\ma{pL}_{\rm m}+\ma{pK}_{\rm m}-2\ma{pH}}e^{-2\phi(d)}-1\right]}{1+10^{\ma{pK}_{\rm m}-
\ma{pH}}e^{-\phi(d)}\left[1+10^{\ma{pL}_{\rm m}-\ma{pH}}e^{-\phi(d)}\right]}\right.\nonumber\\
&&\left.\hspace{3mm}+\frac{1}{1+10^{\ma{pH}-\ma{pT}_{\rm m}}e^{\phi(d)}}\right\}\frac{\sigma_{\rm 0m}}{3}.
\eea

In order to compute the electrostatic potential, we first note that in the acidity regime $2\leq \ma{pH}\leq10$ considered in this work, $\ma{H}^+$ ion density is considerably lower than the 
KCl concentration. Thus,  $\ma{H}^+$ ions will be considered as \textit{spectator ions} that do not contribute to charge screening. Within this approximation, the PB equation reads
\bea\label{pb1}
&&\frac{1}{r}\partial_r\left[r\partial_r\phi(r)\right]-\kappa_{\rm b}^2\sinh\left[\phi(r)\right]\\
&&=-4\pi\ell_{\rm B}\left\{\sigma_{\rm p}\left[\phi(a)\right]\delta(r-a)+\sigma_{\rm m}\left[\phi(d)\right]\delta(r-d)\right\},\nonumber
\eea
with \textcolor{black}{the Bjerrum length $\ell_{\rm B}=e^2/(4\pi\e_{\rm w}k_{\rm B}T)$} and the screening parameter $\kappa_{\rm b}=\sqrt{8\pi\ell_{\rm B}\rho_{\rm b}}$, and the polymer charge density $\sigma_{\rm p}\left[\phi(a)\right]$ 
whose potential dependence will be specified below for the type of polymer under consideration. The integration of Eq.~(\ref{pb1}) around the pore and polymer surface yields the boundary conditions
\be
\label{gau1}
\phi'(a^+)=-4\pi\ell_{\rm B}\sigma_{\rm p};\;\phi'(d^-)=-4\pi\ell_{\rm B}\sigma_{\rm m}.
\ee

To our knowledge Eq.~(\ref{pb1}) cannot be solved in closed form.  \textcolor{black}{Thus, we will solve this equation within an improved Donnan approximation that was introduced in 
Ref.~\cite{Buy2017}. The Donnan approach was shown to be accurate even in the regime of dilute salt $\rho_b=0.01$ M and strong surface charge $\sigma_m=1$ $\mbox{e}/\mbox{nm}^2\approx160$ $\mbox{mC}/\mbox{m}^2$ located well beyond the linearized PB regime.} At the first step, we inject into Eq.~(\ref{pb1}) a uniform Donnan potential ansatz $\phi(r)=\phi_{\rm d}$. 
Integrating the result over the cross-section of the pore, one obtains
\be\label{17}
\sinh(\phi_{\rm d})=\frac{a\sigma_{\rm p}\left[\phi_{\rm d}\right]+d\sigma_{\rm m}\left[\phi_{\rm d}\right]}{\rho_{\rm b}(d^2-a^2)}.
\ee
Equation~(\ref{17}) quartic in the exponential of the potential $\phi_d$ should be solved numerically. Next we improve the pure Donnan approximation 
by taking into account the spatial variation of the potential. To this end, we express the potential as $\phi(r)=\phi_{\rm d}+\delta\phi(r)$ 
and expand Eq.~(\ref{pb1}) at the linear order in the potential correction $\delta\phi(r)$ to get
\be\label{pb2}
\frac{1}{r}\partial_r\left[r\partial_r\phi(r)\right]-\kappa_{\rm d}^2\left[\sinh\left(\phi_{\rm d}\right)+\cosh\left(\phi_{\rm d}\right)\delta\phi(r)\right]=0.
\ee
The solution of Eq.~(\ref{pb2}) reads
\be
\label{don1}
\phi(r)=\phi_{\rm d}-\tanh(\phi_{\rm d})+c_1\ma{I}_0(\kappa_{\rm d} r)+c_2\ma{K}_0(\kappa_{\rm d} r),
\ee
with the pore screening parameter $\kappa_{\rm d}=\kappa_{\rm b} \sqrt{\cosh(\phi_{\rm d})}$ and
\bea
c_1&=&\frac{4\pi\ell_{\rm B}}{\kappa_{\rm d}}\frac{\ma{K}_1(\kappa_{\rm d} a)\sigma_{\rm m}(\phi_{\rm d})+\ma{K}_1(\kappa_{\rm d} d)\sigma_{\rm p}
(\phi_{\rm d})}{\ma{I}_1(\kappa_{\rm d} d)\ma{K}_1(\kappa_{\rm d} a)-\ma{I}_1(\kappa_{\rm d} a)\ma{K}_1(\kappa_{\rm d} d)};\\
c_2&=&\frac{4\pi\ell_{\rm B}}{\kappa_{\rm d}}\frac{\ma{I}_1(\kappa_{\rm d}a)\sigma_{\rm m}(\phi_{\rm d})+
\ma{I}_1(\kappa_{\rm d}d)\sigma_{\rm p}(\phi_{\rm d})}{\ma{I}_1(\kappa_{\rm d}d)\ma{K}_1(\kappa_{\rm d}a)-\ma{I}_1(\kappa_{\rm d}a)\ma{K}_1(\kappa_{\rm d}d)}.
\eea

\section{Results}

\subsection{Apparent zeta potential of the SiN pore}
\begin{figure}
\includegraphics[width=1\linewidth]{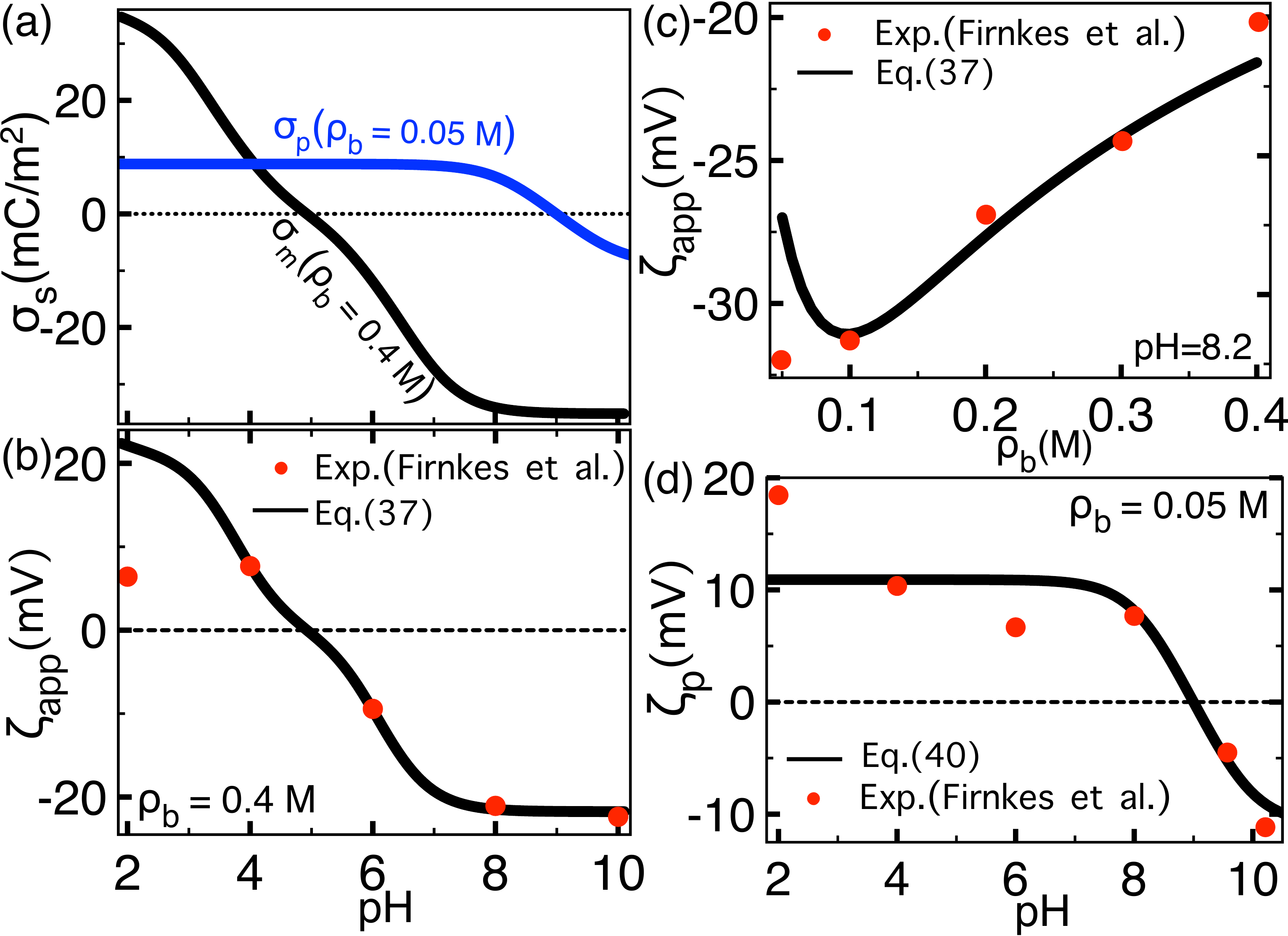}
\caption{(Color online) (a) $\ph$ dependence of the pore (black) and polymer surface charge density (blue). (b) Apparent pore zeta potential~(\ref{ap}) versus the solution 
pH and (c) salt concentration. (d) Polymer zeta potential~(\ref{zep}) against pH. The experimental data in (b) are from Figure 2a of 
Ref.~\cite{Firnkes2010}, the data in (c) from Figure 4 of the supporting information of Ref.~\cite{Firnkes2010}, and the data in (d) from Figure 1b of Ref.~\cite{Firnkes2010}. 
The chemical reaction constants of the pore are $\ma{pK}_{\rm m}=6.1$, $\ma{pL}_{\rm m}=3.75$, and $\ma{pT}_{\rm m}=1.0$, and the dissociable site 
density $\sigma_{\rm 0m}=0.33$ $e/\ma{nm}^2$. The reaction constants for the avidin protein are $\ma{pK}_{\rm p}=9.5$ and $\ma{pL}_{\rm p}=8.5$, 
and the surface density $\sigma_{\rm 0m}=0.055$ $e/\ma{nm}^2$.}
\label{fig2}
\end{figure}

We compare here the experimentally determined apparent zeta potential of the pore obtained from the streaming potential measurements~\cite{Firnkes2010} 
with the theoretical prediction of the present formalism. In the derivation of the apparent zeta potential for the polymer-free pore (i.e. for $a=0$ and $\sigma_{\rm p}=0$), 
we will use the notation of Ref.~\cite{Arel2009}. For a symmetric electrolyte with ionic charges $q_{\pm}=\pm1$ and bulk density $\rho_{\rm b}$, the net charge current through the pore is
\be\label{curr}
I=2\pi e\rho_{\rm b} \sum_{i=\pm}q_i\int_0^d\ma{d}rre^{-q_i\phi(r)}\left[u_{\rm c}(r)+u_{{\rm T}i}\right].
\ee
In Eq.~(\ref{curr}), the convective liquid velocity $u_{\rm c}(r)$ is given by Eq.~(\ref{vels}).  Then, the conductive velocity component of the ionic species $i$ 
reads $u_{{\rm T}i}=-\ma{sign}(q_i)\mu_i\Delta V/L_{\rm m}$, with the mobility of $\ma{K}^+$ ions $\mu_{+}=7.616\times10^{-8}$ $\ma{m}^2\ma{V}^{-1}\ma{s}^{-1}$ 
and $\ma{Cl}^-$ ions $\mu_{-}=7.909\times10^{-8}$ $\ma{m}^2\ma{V}^{-1}\ma{s}^{-1}$~\cite{book}. Substituting into Eq.~(\ref{curr}) the convective and 
conductive velocity components, and using the PB Eq.~(\ref{pb1}), one obtains
\bea
\label{cur2}
I=G_{\rm v} \Delta V+G_{\rm p}\Delta P,
\eea
with the conductance components
\bea
\label{conv}
G_{\rm v}&=&\frac{2\pi e\rho_{\rm b}}{L_{\rm m}}\sum_{i=\pm}\int_0^d\ma{d}rre^{-q_i\phi(r)}\left\{q_i\mu\left[\zeta-\phi(r)\right]-\mu_i|q_i|\right\};\nonumber\\
&&\\
\label{conp}
G_{\rm p}&=&\frac{\pi d^2\mu\zeta}{L_{\rm m}}\left\{\frac{2}{d^2\zeta}\int_0^d\ma{d}rr\phi(r)-1\right\},
\eea
where we introduced the \textcolor{black}{pore} zeta potential $\zeta=\phi(d)$. 

\begin{figure*}
\includegraphics[width=.95\linewidth]{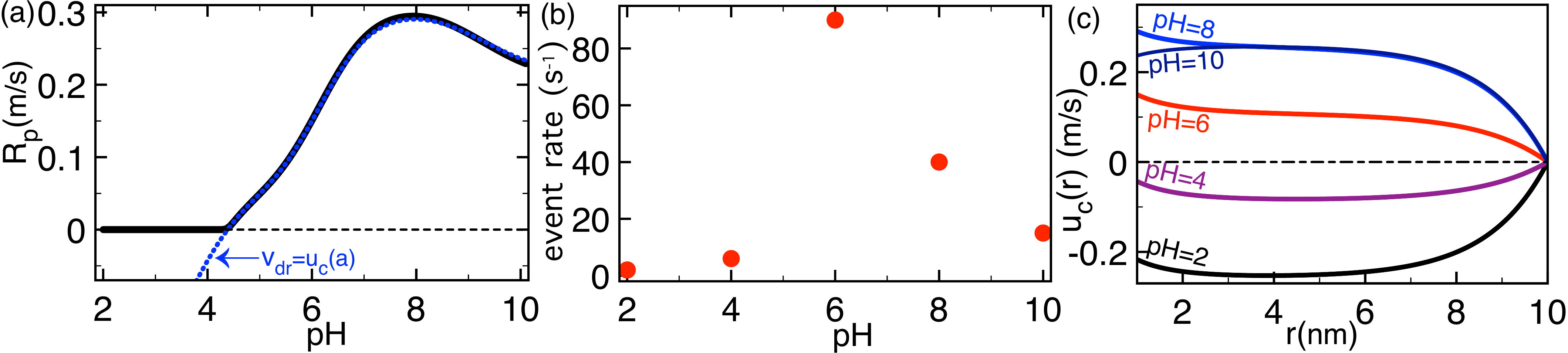}
\caption{(Color online) (a) pH dependence of the avidin translocation rate (black curve) and drift velocity (blue curve), and (b) 
experimental rate of translocation events from TABLE 1 of Ref.~\cite{Firnkes2010}. (c) Liquid velocity profile~(\ref{vels}) at various pH. The salt concentration 
$\rho_{\rm b}=0.05$ M and external voltage $\Delta V=-150$ \textcolor{black}{m}V are taken from of Ref.~\cite{Firnkes2010}. In the corresponding drift regime, the curves have no 
visible dependence on the precise value of the polymer length set to $L_{\rm p}=10$ nm.}
\label{fig3}
\end{figure*}

The streaming potential corresponds to the voltage that cancels the current~(\ref{cur2}), i.e. $\Delta V_{\rm str}=-\left(G_{\rm p}/G_{\rm v}\right)\Delta P$. Introducing the reduced conductivities
\bea
\label{con1}
K_{\rm v}&=&\frac{2}{d^2}\left\{\sum_{i=\pm}\frac{\sigma_i}{\sigma_{\rm T}}\int_0^d\mathrm{d}rr\left[e^{-q_i\phi(r)}-1\right]\right.\\
\label{con2}
&&\left.\hspace{8mm}+\frac{\mu e}{\sigma_{\rm T}}\int_0^d\mathrm{d}rr\sum_{i=\pm}q_ie^{-q_i\phi(r)}\left[\phi(r)-\zeta\right]\right\};\nonumber\\
K_{\rm p}&=&\frac{2}{d^2\zeta}\int_0^d\ma{d}rr\phi(r),
\eea
with the bulk conductivity of the species $i$ $\sigma_i=e\mu_i|q_i|\rho_{{\rm b}i}$ and the total conductivity $\sigma_{\rm T}=\sigma_+ + \sigma_-$, one obtains
\be
\Delta V_{\rm str}=-\frac{\e_{\rm w}k_{\rm B}T\zeta_{\rm app}}{e\eta\sigma_{\rm T}}\Delta P,
\ee
where the apparent zeta potential is given by
\be\label{ap}
\zeta_{\rm app}=\frac{1-K_{\rm p}}{1+K_{\rm v}}\zeta.
\ee
At the bulk KCl concentration $\rho_{\rm b}=0.4$ M, our computed bulk conductivity $\sigma_{\rm T}=6.0$ S/m compares well with the 
experimentally measured value of $4.7-5.1$ S/m\cite{Firnkes2010}.

Figures~\ref{fig2}(a) and (b) display the pH dependence of the surface charge and apparent zeta potential of the SiN pore~\cite{Firnkes2010}. The chemical parameters 
providing the best agreement with the experimentally measured zeta potential are given in the legend. Starting at $\ma{pH}=10$ and rising the acidity of the solution, 
$\ma{H}^+$ binding to the silanol and primary amine groups increases the pore charge and zeta potential ($\ma{pH}\downarrow\sigma_{\rm m}\uparrow\zeta_{\rm app}\uparrow$), 
and turns them from negative to positive at $\ma{pH}\approx5$. Our model can accurately reproduce the pH dependence of the experimental data, except at 
$\ma{pH}=2$ where the data is overestimated.

Figure~\ref{fig2}(c) displays the salt dependence of the apparent zeta potential $\zeta_{\rm app}$ at $\ma{pH}=8.2$ where the pore is anionic. One sees that salt addition 
amplifies charge screening and lowers the magnitude of this potential, i.e. $\rho_{\rm b}\uparrow|\zeta_{\rm app}|\downarrow$. With the same model parameters as in 
Fig.~\ref{fig2}(b), the theoretical prediction for $\zeta_{\rm app}$  agrees well with the experimental data. As the apparent zeta potential~(\ref{ap}) involves, in addition to the 
bare potential $\zeta$, the pore conductance components~(\ref{con1}) and~(\ref{con2}), the agreement with experiments indicates that our model is also accurate in  
predicting the pressure and voltage-driven charge conductivity of the pore. \textcolor{black}{In order to identify the cause of the deviation between theory and experiments in the low pH and salt density regimes, comparisons with additional experimental electrokinetic data and the relaxation of the model approximations introduced in our work will be needed.}

\subsection{Voltage-driven translocation of avidin proteins}
\label{av}

We investigate here the pH controlled translocation of avidin proteins through SiN nanopores under an externally applied voltage~\cite{Firnkes2010}. 
According to the zeta potential measurements of Ref.~\cite{Firnkes2010}, avidin is an amphoteric polyelectrolyte. Thus, we model the pH dependent inversion 
of the avidin charge by the chemical reaction scheme
\be\label{13}
\m{SOH}\ra[k'_{\rm r}]{k'_{\rm d}}\m{SO}^-+\m{H}^+;\;\m{SOH}+\m{H}^+\ra[l'_{\rm d}]{l'_{\rm r}}\m{SOH}_2^+,
\ee
with the characteristic dissociation rates $K_{\rm p}=10^{-\ma{pK}_{\rm p}}=k'_{\rm d}/k'_{\rm r}$ and $L_{\rm p}=10^{-\ma{pL}_{\rm p}}=l'_{\rm d}/l'_{\rm r}$. 
Following the derivation of Eq.~(\ref{11}), the protein charge density follows as
\be
\label{14}
\sigma_{\rm p}=\frac{10^{\ma{pL}_{\rm p}+\ma{pK}_{\rm p}-2\ma{pH}}e^{-2\phi(a)}-1}{1+10^{\ma{pK}_{\rm p}-\ma{pH}}e^{-\phi(a)}\left[1+10^{\ma{pL}_{\rm p}-
\ma{pH}}e^{-\phi(a)}\right]}\sigma_{\rm 0p},
\ee
where $\sigma_{\rm 0p}$ stands for the density of the dissociable groups.  

\begin{figure*}
\includegraphics[width=1\linewidth]{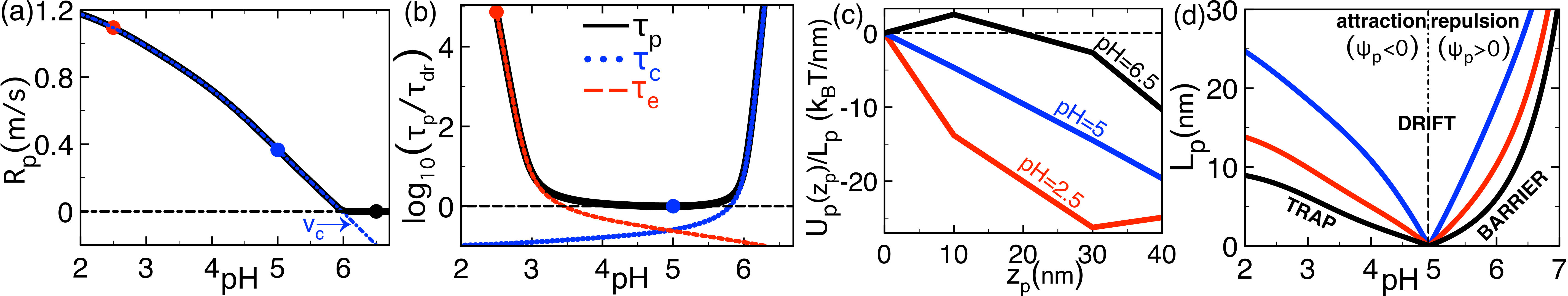}
\caption{(Color online) $\ph$ dependence of (a) the ds-DNA translocation rate (black curve) and capture velocity~(\ref{15}) (blue curve), and 
(b) translocation time, capture time, and escape time rescaled by the drift limit $\tau_{\rm dr}$. (c) Polymer potential~(\ref{polp2}) at various $\ph$. 
In (a)--(c), the salt density is $\rho_{\rm b}=0.005$ M. (d) $\ph$ dependence of the critical polymer length~(\ref{lcr}) separating the polymer trapping, barrier, and drift transport 
regimes at the salt densities $\rho_{\rm b}=0.001$ (blue), $0.005$ (red), and $0.01$ M (black). The voltage is $\Delta V=100$ mV in all plots. The remaining parameters are the same as in Fig.~\ref{fig3}.}
\label{fig4}
\end{figure*}

Figure~\ref{fig2}(d) compares the avidin zeta potential obtained from the charge regulation scheme of Eq.~(\ref{13}) with the experimental values of 
Ref.~\cite{Firnkes2010} extracted from the polymer mobility. The theoretical prediction for the zeta potential is obtained from the bulk limit of 
Eq.~(\ref{don1}) where $\phi_{\rm d} \to0$ and $\kappa_{\rm d} \to\kappa_{\rm b}$, which yields
\be\label{zep}
\zeta_{\rm p}=\lim_{d\to\infty}\phi(a)=\frac{4\pi\ell_{\rm B}\sigma_{\rm p}(0)}{\kappa_{\rm b}}\frac{\mathrm{K}_0(\kappa_{\rm b}a)}{\mathrm{K}_1(\kappa_{\rm b}a)}.
\ee
The chemical reaction parameters providing the best agreement with the experimental data are given in the legend of Fig.~\ref{fig2}. One notes that the 
pH reduction increases the avidin charge (the blue curve in Fig.~\ref{fig2}(a)) and the zeta potential ($\ma{pH}\downarrow\sigma_{\rm p}\uparrow\zeta_{\rm p}\uparrow$), 
and switches their sign at the point of zero charge $\ma{pH}\approx9$. Within the experimental scattering, the charge regulation model~(\ref{13}) 
can account for the pH induced inversion of the avidin zeta potential with a reasonable accuracy. 

Having established the $\ph$ dependence of the pore and protein surface charges, we characterize the avidin conductivity of the SiN pore. 
Figures~\ref{fig3}(a) and (b) display respectively the translocation rate \textcolor{black}{in Eq.~(\ref{rc})} and the experimental rates of translocation events from Ref.~\cite{Firnkes2010}. 
One notes that for $\ma{pH}\lesssim4$, translocation events are rare. At $\ma{pH}\gtrsim4$, the translocation rate quickly rises ($\ph\uparrow R_{\rm p}\uparrow$), 
reaches a peak at $\ma{pH}\sim6-8$, and drops beyond this value ($\ph\uparrow R_p\downarrow$). The comparison of Figs.~\ref{fig3}(a) and (b) 
shows that  our model can accurately reproduce the overall pH dependence of the experimental translocation data.  
The slower decay of the theoretical curve at large $\ph$ may be due to the contribution from the diffusion-limited capture regime not included in our model.

According to Eqs.~(\ref{rc}) and~(\ref{polp}),  translocation is driven by electrostatic polymer-pore interactions embodied in the potential $V_{\rm p}(z_{\rm p})$, 
and the EP and EO drags resulting in the velocity $v_{\rm dr}$. In Fig.~\ref{fig3}(a), the strong correlation between the $v_{\rm dr}$ and $R_{\rm p}$ 
curves implies that due to the weak avidin surface charge, avidin translocation is drift-driven and protein-pore interactions play a minor role. 

To characterize the $\ph$ dependence of the avidin translocation rates in terms of the electrohydrodynamic drift, in Fig.~\ref{fig3}(c) we report the liquid velocity 
profile~(\ref{vels}) at various pH values. This plot should be interpreted together with the surface charge density plots in Fig.~\ref{fig2}(a). 
We note that the electric field $E$ induced by the negative voltage $\Delta V=-150$ mV is oriented towards the trans side corresponding to the positive velocity direction 
(see Fig.~\ref{fig1}). From $\ph=2$ to $4$, the $\ma{Cl}^-$ ions attracted by the cationic pore ($\sigma_{\rm m}>0$) result in a negative liquid velocity $u_{\rm c}(r)<0$. 
As $\sigma_{\rm m}>\sigma_{\rm p}$, the corresponding EO drag in the cis direction dominates the EP drift in the trans direction and results in a negative polymer 
velocity $v_{\rm dr}=u_{\rm c}(a)<0$. Thus, the hinderance of polymer capture at $\ph\leq4$ stems from the drag force induced by the anionic EO flow.

Rising the solution $\ph$ in the subsequent regime $4\leq\ph\leq8$, the protein charge $\sigma_{\rm p}$ remains constant while the pore charge 
$\sigma_{\rm m}$ drops and turns from positive to negative. The resulting cation excess leads to a positive EO velocity $u_{\rm c}(r)>0$ and polymer drift 
velocity $v_{\rm dr}=u_{\rm c}(a)>0$ (see Fig.~\ref{fig3}(c)). Thus, the quick rise of the event rates at $\ph>4$  is induced by the cationic EO flow that 
drags the protein in the trans direction. Finally, increasing the $\ph$ beyond the value $\ph\sim8$, $\sigma_{\rm m}$ remains constant while $\sigma_{\rm p}$ 
turns from positive to negative. The protein charge inversion switches the sign of the avidin zeta potential $\phi(a)$ and turns the direction of the EP velocity component 
$v_{\rm ep}=\mu E\phi(a)$ from trans to cis side, reducing the translocation rate in Figs.~\ref{fig3}(a) and (b). Thus, beyond the charge inversion point 
$\ph\approx9$, protein capture is solely driven by the EO flow. These results confirm a similar mechanism that was proposed in  
Ref.~\cite{Firnkes2010} based on the comparison of the experimental pore and protein zeta potentials.

\subsection{pH controlled DNA trapping}
\label{dna}

In nanopore-based biosensing approaches, serial polymer translocation necessitates fast polymer capture while accurate sequencing requires  long signal duration, i.e. 
extended translocation time. We characterize the ds-DNA conductivity of SiN pores to show that mutual enhancement of the polymer capture speed and translocation time 
can be achieved by tuning the acidity of the solution. We have recently showed that ds-DNA transport can be accurately described by an inert polymer surface 
charge~\cite{Buy2018}. Thus, we fixe the DNA surface charge density to the value $\sigma_{\rm p}=-0.4$ $\ma{e}/\ma{nm}^2$ obtained from current blockade data~\cite{Buy2014}.

Figures~\ref{fig4}(a)--(c) display the $\ph$ dependence of the ds-DNA translocation rate $R_{\rm p}$ and rescaled translocation time $\tau_{\rm p}/\tau_{\rm dr}$ (black curves), 
and the polymer potential profile $U_{\rm p}(z_{\rm p})$. The behavior of these quantities can be described in terms of the inverse lengths \textcolor{black}{$\lambda_{\rm d}$ and $\lambda_{\rm b}$} introduced in 
Eq.~(\ref{len}). At $\ph=6.5$ where the system is located in the barrier-dominated regime $\lb>\ld$, the pore entrance is characterized by an electrostatic barrier that reaches the value 
$\textcolor{black}{U_{\rm p}/L_p} \approx2.5$ $k_{\rm B}T/\ma{nm}$  at $z_{\rm p}=L_{\rm p}=10$ nm. Figure~\ref{fig4}(a) shows that this barrier leads to a negative capture velocity
\be
\label{15}
v_{\rm c}=v_{\rm p}(z_{\rm p}<L_{\rm p})=v_{\rm dr}\left(1-\frac{\lb}{\ld}\right),
\ee
resulting in a vanishingly small translocation rate $R_{\rm p}$ and large translocation time $\tau_{\rm p}$. Thus, at large $\ph$ values where the membrane is anionic, 
polymer capture is hindered by electrostatic DNA-pore repulsion. Then, the increase of the acidity to the point of zero charge $\ph=5$ suppresses the barrier and takes the 
system to the drift-driven regime $\ld>\lb>-\ld$ where the polymer potential $U_{\rm p}(z_{\rm p})$ turns to downhill. This enhances the capture velocity and translocation rate, 
and reduces the translocation time ($\ph \downarrow v_{\rm c} \uparrow R_{\rm p} \uparrow\tau_{\rm p} \downarrow$) by orders of magnitude. 

Below the value $\ph\approx5$ where the pore becomes cationic, the translocation rate and time rise mutually with the acidity of the solution, i.e. 
$\ph \downarrow R_{\rm p} \uparrow \tau_{\rm p} \uparrow$. This departure from the drift transport picture originates from the onset of opposite charge 
DNA-pore interactions.  Indeed, Fig.~\ref{fig4}(c) shows that the variation of the acidity from $\ph=5$ to $2.5$  lowers the potential $U_{\rm p}(z_{\rm p})$ 
and gives rise to an attractive potential minimum at the pore exit $z_{\rm p}=L_{\rm m}=30$ nm \textcolor{black}{(see also the top plot in the inset of Fig.~\ref{fig1})}. At the corresponding $\ph$ value,  the system is located in the trapping regime 
$\lb<-\ld$ where the polymer-pore attraction enhances the DNA capture velocity~(\ref{15})  ($v_{\rm c}>v_{\rm dr}$) but also traps the molecule at the pore exit. 
Figure~\ref{fig4}(b) shows that upon the variation of the acidity from $\ph=6.5$ to $2.5$, this mechanism reduces the polymer capture time and increases the polymer escape time 
($\ph\downarrow\tau_{\rm c}\downarrow\tau_{\rm e}\uparrow$) from their drift limit by several orders of magnitude. 
This prediction is of high relevance for the optimization of nanopore-based biosensing techniques.

The effect of the polymer length on these features can be characterized by recasting the capture velocity~(\ref{15}) as
\be
\label{16}
v_{\rm c}=v_{\rm dr}\left[1-\ma{sign}(\psi_{\rm p})\frac{L_{\rm p}^*}{L_{\rm p}}\right],
\ee
with the critical length
\be\label{lcr}
L_{\rm p}^*=\frac{\ln(d/a)\left|\psi_{\rm p}\right|}{2\pi\eta\beta v_{\rm dr}}
\ee
separating the drift ($L_{\rm p}>L_{\rm p}^*$) and barrier/trapping regimes ($L_{\rm p}<L_{\rm p}^*$). Figure~\ref{fig4}(d) displays the $\ph$ dependence of the length~(\ref{lcr}). The location of the barrier and trapping regimes below the critical line  $L_{\rm p}^*-\ph$ stems from the fact that the external voltage acts on the whole polymer sequence while polymer-pore interactions originate solely from the 
polymer portion in the pore. Thus, polymer-pore interactions have a stronger effect on the translocation of shorter sequences. According to Eq.~(\ref{16}), this 
results in the faster capture of shorter polymers by cationic pores, i.e. $L_{\rm p}\downarrow v_{\rm c}\uparrow$ for $\psi_{\rm p}<0$. 
One also notes that in the same cationic pore regime of Fig.~\ref{fig4}(d), due to the enhancement of the polymer-pore attraction, 
the upper length~(\ref{lcr}) for polymer trapping rises with increasing acidity ($\ph \downarrow L_{\rm p}^*\uparrow$) and decreasing salt 
($\rho_{\rm b} \downarrow L_{\rm p}^*\uparrow$).  This phase diagram may provide guiding information for probing the 
$\ph$ controlled polymer trapping in translocation experiments.

\section{Summary and Conclusions}

In this Letter we have introduced an electrohydrodynamic model of $\ph$ controlled polymer translocation through SiN pores whose surface charge can be inverted upon protonation. 
Our model incorporates a charging procedure that can quantitatively reproduce the experimentally established $\ph$ and salt dependence of the pore surface charge. 
Within the framework of this model, we have investigated the electrohydrodynamic mechanism behind the avidin conductance of SiN pores. 
Our model can accurately reproduce the experimentally measured non-monotonic dependence of the avidin translocation rates on the solution 
$\ph$~\cite{Firnkes2010}. We showed that this peculiarity originates from the interplay between the EO drag and EP drift forces on the avidin protein.

We have also investigated the transport of ds-DNA molecules through SiN pores. Our analysis unraveled an electrostatic trapping mechanism that allows the mutual increase 
of the polymer capture speed and translocation time by $\ph$ tuning. As polymer trapping occurs in the escape regime $z_{\rm p}>\lm$, the scanning of the entire 
polymer sequence at reduced velocity is possible only if the pore is longer than the polymer. Our finite-size analysis also shows that faster polymer 
capture followed by extended translocation occurs for sequences of length $L_{\rm p}<L_{\rm p}^*$. This inequality is consistent with the above-mentioned length 
hierarchy $L_{\rm p}<L_{\rm m}$ required for the slow sequencing of the entire polymer in the electrostatic trap. We have also shown that the upper sequence length 
$L_{\rm p}^*$ for polymer trapping can be tuned upon the variation of the acidity or the salt concentration. \textcolor{black}{Future works can extend our model by accounting for ion and solvent specific effects, more sophisticated charging models, the diffusion-limited capture regime, the electrostatics of the finite membrane size, and entropic polymer fluctuations.}
\\
\acknowledgements  This work was performed as part of the Academy of Finland QTF Centre of Excellence program (project 312298) and has also been supported
by the Aalto Science Institute through a sabbatical grant (S.B.).

\end{document}